\documentclass{aastex631}

\received{}
\revised{}
\accepted{}

\usepackage[utf8]{inputenc}
\usepackage{graphicx}
\usepackage{booktabs}
\usepackage{geometry}
\usepackage{color}
\usepackage{bm}
\usepackage{cases}
\usepackage{multirow}
\usepackage{lineno}
\shorttitle{Calibrating GRBs by using Gaussian Process with SN Ia}
\shortauthors{Liang et al.}

\begin{document}

\title{Calibrating Gamma-Ray Bursts by Using a Gaussian Process with Type Ia Supernovae}

\author{Nan Liang}\thanks{liangn@bnu.edu.cn}
\affiliation{Key Laboratory of Information and Computing Science Guizhou Province, Guizhou Normal University, Guiyang, Guizhou 550025, China}
\affiliation{Joint Center for FAST Sciences Guizhou Normal University Node, Guiyang, Guizhou 550025, China}

\author{Zihao Li}\thanks{lizihao@gznu.edu.cn}
\affiliation{Key Laboratory of Information and Computing Science Guizhou Province, Guizhou Normal University, Guiyang, Guizhou 550025, China}
\affiliation{Joint Center for FAST Sciences Guizhou Normal University Node, Guiyang, Guizhou 550025, China}

\author{Xiaoyao Xie} \thanks{xyx@gznu.edu.cn}
\affiliation{Key Laboratory of Information and Computing Science Guizhou Province, Guizhou Normal University, Guiyang, Guizhou 550025, China}
\affiliation{Joint Center for FAST Sciences Guizhou Normal University Node, Guiyang, Guizhou 550025, China}

\author{Puxun Wu}\thanks{pxwu@hunnu.edu.cn}
\affiliation{Department of Physics and Synergistic Innovation Center for Quantum Effects and Applications, Hunan Normal University, Changsha, Hunan 410081, China}


\begin{abstract}
    In this paper, we calibrate the Amati relation  (the $E_{\rm p}$-${E}_{\rm iso}$ correlation) of gamma-ray bursts (GRBs) in a cosmology-independent way. By using Gaussian process to reconstruct the smoothed luminosity distance from the Pantheon type Ia supernovae (SNe Ia) sample, we utilize the reconstructed results to calibrate the $E_{\rm p}$-${E}_{\rm iso}$ correlation with the Markov Chain Monte Carlo method and construct a Hubble diagram with the A220 GRB data, in which there are A118 GRB data with the higher qualities appropriate for cosmological purposes. With 98 GRBs at $1.4<z\leq8.2$ in the A118 sample and the observed Hubble data, we obtain $\Omega_{\rm m}$=$0.346^{+0.048}_{-0.069}$, $h$=$0.677^{+0.029}_{-0.029}$  for the flat $\Lambda$CDM model, and  $\Omega_{\rm m}$=$0.314^{+0.072}_{-0.055}$, $h$=$0.705^{+0.055}_{-0.069}$, $w$=$-1.23^{+0.33}_{-0.64}$ for the flat $w$CDM model, which are consistent with those from fitting the coefficients of the Amati relation  and the cosmological parameters simultaneously.
   \end{abstract}

\section{Introduction}
Observations of type Ia supernovae (SNe Ia) provide a powerful probe in modern cosmology, from which  the accelerated expansion of the universe has been found \citep{Riess1998,Perlmutter1999}.  The maximum redshift observed by SNe Ia  is about $z\sim2.3$ \citep{Scolnic2018}. Thus,  to explore the cosmic evolution at the high-redshift region requires  observing more luminous objects than SNe Ia. Gamma-ray bursts (GRBs) are the strongest bursts of high-energy gamma rays from cosmological space in a short time, which are the most intense explosions
observed so far. At present, the maximum redshift of the GRB can reach at $z=9.4$ \citep{Cucchiara2011}.  Therefore, GRBs can be used to probe the universe at the high redshift.
Utilizing GRB in cosmology requires its luminosity relations, which are connections between measurable properties of the instantaneous gamma-ray emission and the luminosity or energy. Several empirical GRB luminosity relations have been proposed \citep{Fenimore2000,Norris2000,Amati2002,Ghirlanda2004a,Yonetoku2004,Liang2005,Firmani2006,Dainotti2008,Yu2009,Tsutsui2009a,Izzo2015}, see e.g., \cite{Ghirlanda2006}, and \cite{Schaefer2007} for reviews.
According to these relations, the GRBs have  been used as the cosmic probe   for  researching the evolving history of our universe and the properties of dark energy \citep{Ghirlanda2004b,Dai2004,Firmani2005,Xu2005,Liang2006,Wang2006,Schaefer2007}.
For recent reviews of GRB luminosity relations and their applications in cosmology, see e.g.  \cite{Wang2015}, \cite{Dainotti2017} and \cite{Dainotti2018}.

In the early cosmological research works of GRBs, a certain cosmological model was assumed to calibrate the GRB luminosity relation \citep{Schaefer2003,Dai2004,Schaefer2007}. When using these model-dependent GRB data to constrain the cosmological model, it  suffers the so-called circularity problem \citep{Ghirlanda2006}. In order to avoid this circularity problem in the application of GRBs in cosmology,  the simultaneous fitting method  has been proposed \citep{Amati2008,Li2008,Wang2008}, in which the coefficients of relations and the parameters of the cosmological model are constrained simultaneously.
However, the circularity problem cannot be circumvented completely by means of statistical approaches, because a particular cosmological model is required in doing the joint fitting.
Recently, \cite{Khadka2020} fitted simultaneously the cosmological and GRB relation parameters in a number of different
cosmological models, and found that the Amati relation parameters are almost identical in all cosmological models, which seems to indicate that these GRB data sets are standardizable within the error bars.

On the other hand, \cite{Liang2008} proposed a cosmological model-independent method to calibrate the luminosity relations of GRBs by using the SNe Ia data. It is obvious that objects at the same redshift should have the same luminosity distance in any cosmology. Therefore, in the same sense as using Cepheid variables to calibrate SNe Ia, if regarding SNe Ia as the first-order standard candles, GRBs can be calibrated from SNe Ia in a completely cosmological model-independent way. The luminosity distances at the redshift of the low-redshift GRB data can be derived by interpolating the SNe Ia data directly, and then the values of the coefficients of the GRB luminosity relation can be obtained from these low-redshift GRB data. Extrapolating these results on the high-redshift GRB data can build the GRB Hubble diagram. Thus, the standard Hubble diagram method can be used to constrain the cosmological model \citep{Capozziello2008,Capozziello2009,Izzo2009,Wei2009,Wei2010,Liang2010,Liang2011,Freitas2011,Wang2011,Wei2015,Wang2016}.
Similar to the interpolation method, GRBs are calibrated
from the SNe Ia by using the polynomial fitting \citep{Kodama2008,Tsutsui2009b}, an iterative procedure \citep{LiangZhang2008}, the local regression \citep{Cardone2009,Cardone2011,DP2011,Demianski2011,Demianski2017a}, the cosmography methods \footnote{Model-independent techniques to calibrate GRB correlations have been investigated in the field of cosmography (see, e.g. \cite{Luongo2020}). The results look similar to those here presented, which are reasonable and interesting. Moreover, the cosmographic technique has been wildly used by several approaches in pure cosmology to investigate the expansion history of the universe (see, e.g. \cite{Aviles2012,Gruber2014,Dunsby2016,Capozziello2018}
).}\citep{Capozziello2010,WangDai2011,Gao2012,Wang2014}, a two-steps method
minimizing the use of SNe Ia \citep{Izzo2015,Muccino2021}, and the Pad\'e approximation method \citep{Liu2015}.

A Gaussian process is a fully Bayesian approach for smoothing data, which can effectively reduce the errors of reconstructed results compared to the approaches mentioned in the above.
In recent years, a Gaussian process method has been widely applied to the field of cosmology and astrophysics \citep{Seikel2012a,Seikel2012b,Busti2014,Yu2016,Yu2018,Lin2018,Wei2018,Pan2020,Sun2021,Avila2022}. For examples,  \cite{Seikel2012a} used a Gaussian process to reconstruct the luminosity distance with its derivatives and the dark energy dynamics from SNe Ia.  \cite{Lin2018} constrained the distance duality relation with the Gaussian process from SNe Ia, galaxy clusters, and baryon acoustic oscillations.
\cite{Sun2021} investigated the influence of the bounds of the hyperparameters on the reconstruction of the Hubble constant with the Gaussian process.

The Amati relation \citep{Amati2002}, which connects the spectral peak energy and the isotropic equivalent radiated energy (the $E_{\rm p}$-${E}_{\rm iso}$ correlation) of GRBs, has been widely used in GRB cosmology \citep{Amati2008,Wei2009,DP2011,Demianski2011,Demianski2017a,Demianski2017b,Liu2015,Feng2016}. \cite{Wei2010} calibrated 109 GRBs with the Amati relation, using the cosmology-independent calibration method proposed by \cite{Liang2008}.\cite{Wang2016} used two model-independent methods to standardize the Amati relation with 151 GRB data (including the update 42 GRBs).
Recently, \cite{Amati2019} proposed another similar cosmological model-independent method to calibrate the Amati relation  by using the observed Hubble data (OHD) through the B\'ezier polynomial, and built up a new data set consisting of 193 GRBs (with firmly measured redshift and spectral parameters taken from \cite{Demianski2017a} and references therein). \cite{Dirirsa2019} found that the Amati relation is satisfied by the 25 Fermi GRB sample. For comparisons, the authors also use a sample of 94 GRBs selected from 151 GRBs analyzed by  \cite{Wang2016}. For recent works that used the Amati relation for the application in cosmology, see, e.g. \cite{Wang2019,Khadka2020,Shirokov2020,Demianski2021,Montiel2021,Tang2021,Luongo2021a,Luongo2021b,Khadka2021,Cao2021,CaoRatra2022,Gowri2022,Jia2022},
and \cite{Muccino2022}.

More recently, \cite{Khadka2021} used the Amati relation and the Combo-correlated GRB data sets\footnote{The Combo relation \citep{Izzo2015} is an hybrid correlation involving prompt and afterglow GRB parameters (i.e., the plateau luminosity $L_0$, the rest-frame duration $\tau$, and the late power-law decay index $\alpha$) with a small data scatter, which has been investigated by \cite{Muccino2021,Luongo2021a} and \cite{Tang2022}.}  to simultaneously derive the correlation and cosmological model parameter
constraints. For the Amati relation, the authors compile a data set of 118 bursts (the A118 sample), including recent Fermi observations samples  from the total 220 GRBs (the A220 sample) with the smallest intrinsic dispersion, which is  suitable for constraining cosmological parameters. With the A220 and the A118 GRB samples,
\cite{Cao2022a,Cao2022b} have used the Amati relation in conjunction with the Dainotti-correlated GRB data
sets\footnote{The Dainotti relation  \citep{Dainotti2008} between the plateau luminosity ($L_0$) and the end time of the plateau in X-ray afterglows ($t_b$) have been used for cosmological purposes \citep{Hu2021,Wang2022,Cao2022a,Cao2022b,Dainotti2022a,Dainotti2022b,Dainotti2022c}.
The similar relations with the plateau in the X-ray afterglows have been used in \cite{Xu2021}.}
 compiled recently by \cite{Hu2021} and \cite{Wang2022} to
constrain cosmological model parameters;
\cite{Liu2022a, Liu2022b} have  proposed the improved Amati relations, which contains a redshift-dependent term,  via a powerful statistical tool called copula.

In this paper, we plan to use the Gaussian process to reconstruct the luminosity distance from the Pantheon SNe Ia sample \citep{Scolnic2018}, without assuming any specific form of the distance-redshift relation of SNe Ia, and then calibrate the Amati relation with the total 220 GRB samples and the A118 sample \citep{Khadka2021} to obtain the GRB Hubble diagram at the high redshift.
With 98 GRB data  at $1.4<z\leq8.2$ in the A118 sample and the OHD, we constrain the $\Lambda$CDM model and $w$CDM model in flat space. Finally, we also use GRB data sets to constrain the cosmological models and GRB relation parameters simultaneously.

\section{GRB HUBBLE DIAGRAM FROM LOW-REDSHIFT CALIBRATION}

\subsection{GRBs calibration with the Amati relation at $z<1.4$}

The Gaussian process can reconstruct effectively a smooth function from the discrete data points without assuming  explicit fitting forms of the function.  In the Gaussian process, the reconstructed function is a Gaussian random variable at a reconstructed point, which is completely confirmed by its mean function and covariance function. The function values $f(z)$  are correlated by a covariance function $k(z,\tilde z)$ to characterize the connection between the function values at different reconstructed points \citep{Seikel2012a}. There are a lot of covariance functions available that we can choose. The advantage of the squared exponential covariance function is that it is infinitely differentiable, which is useful for reconstructing the derivative of a function \citep{Seikel2012a}. The squared exponential covariance function is given by
\begin{equation}
  k(z,\tilde{z})=\sigma_f^2\exp\left[-\frac{(z-\tilde{z})^2}{2l^2}\right].
\end{equation}
 The hyperparameter $\sigma_f$, which determines the typical change of $f(z)$, and $l$, which determines the length in the $z$-direction, can be optimized by maximizing the marginal likelihood.

We use public python package \texttt{GaPP}\footnote{\url{https://github.com/astrobengaly/GaPP}} to calibrate the GRB relation from the SNe Ia. For the GRB data set, we use the total 220 GRB data (A220)\footnote{The A220 data set is composed of 220 long GRBs \citep{Khadka2021}, including A118 data sets, as well as 102 data sets (A102) from 193 GRBs analyzed by \cite{Amati2019} and \cite{Demianski2017a}, which have not already been included in the A118 sample.} including the recent Fermi observations \citep{Khadka2021}, as well as the higher-quality 118 data set (A118)\footnote{The A118 data set is composed of 118 long GRBs \citep{Khadka2021}, including 25 long GRBs with Fermi-GBM/LAT data and well-constrained spectral properties \citep{Dirirsa2019}, as well as 93 bursts updated from a sample of 94 GRBs  (with GRB 020127 removed because its redshift is not secure) selected from 151 GRBs analyzed by \cite{Wang2016}.} with a tighter intrinsic scatter. For SNe Ia data sets, we use the Pantheon sample \citep{Scolnic2018}, which contained 1048 SNe Ia data points with the apparent magnitude. The distance modulus relates to the luminosity distance $d_L$ through
$\mu = m - M = 5{\rm log}_{10}(\frac{d_{L}}{{{\rm Mpc}}}) + 25$, where
$m$ and $M$ are the apparent magnitude and the absolute magnitude, respectively.
In this procedure, we reconstruct the apparent magnitude of GRBs from SNe Ia. The apparent magnitudes reconstructed from the Gaussian process
with the 1$\sigma$ uncertainty from SNe Ia data are plotted in Figure 1.

We find that the reconstructed function presents strange oscillations with a large uncertainty in the range where data points are sparse at $1.4\leq z\leq 2.3$. Thus,  the lack of SNe Ia at $z\geq1.4$ in the \texttt{GaPP} procedure will produce the limits of Gaussian processes, which may affect the overall analysis of data comparisons. However, after removing SNe Ia data at $1.4\leq z\leq2.3$, we find that the apparent magnitudes  reconstructed from SNe Ia data at $z<1.4$ are almost identical with those reconstructed from SNe Ia data at $z\leq2.3$.  This indicates that this limitation does not affect the reconstructed results in the redshift region $z<1.4$,  which can be seen in Figure 1. Therefore we can use  the luminosity distance reconstructing from SNe Ia at $z<2.3$ to calibrate the Amati relation with 79 GRBs of A220 and 20 GRBs of A118 at $z<1.4$.
\begin{figure*}[h]
\centering
\includegraphics[width=360px]{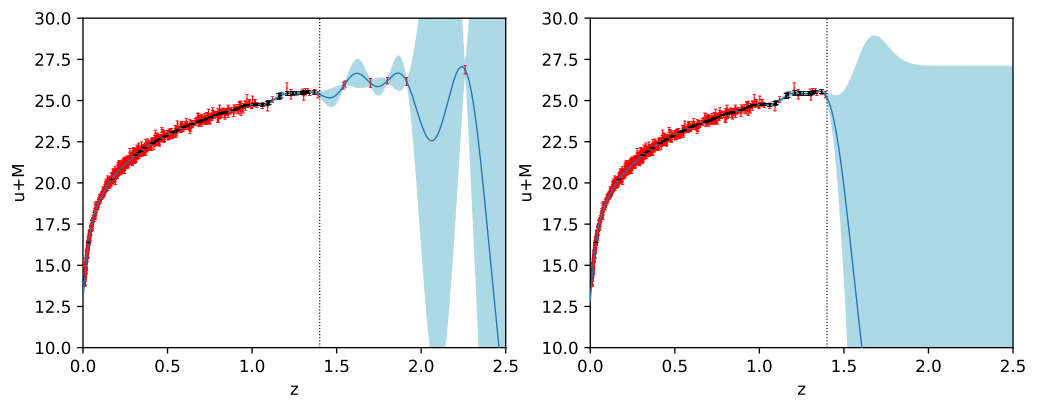}
\caption{The apparent magnitudes reconstructed through the Gaussian process 
from  SNe Ia data at $z\leq2.3$ (left panel), and those reconstructed from SNe Ia data at $z<1.4$ (right panel). The blue curves present the reconstructed function with the 1$\sigma$ uncertainty  from the SNe Ia data (red dots). The apparent magnitudes of GRBs at $z<1.4$ (black dots) are reconstructed from SNe Ia through the Gaussian process. The dashed line denotes $z=1.4$.} \label{The apparent magnitudes reconstructed from  Gaussian processes}
\end{figure*}


The Amati relation can be expressed as
\begin{equation}y= a + bx,\end{equation}
where $y={\rm log}_{10}\frac{E_{{\rm is o}}}{{\rm 1erg}}$, $x={\rm log}_{10}\frac{E_{{\rm p}}}{{\rm 300keV}}$, $E_{\rm p}$ and ${E}_{\rm iso}$ are the spectral peak energy and the isotropic equivalent radiated energy,
and $a$ and $b$ are free coefficients needing to be calibrated from  the observed data. $E_{{\rm iso}}$ and $E_{\rm p}$ can be calculated through
\begin{equation}E_{{\rm iso}} = 4\pi d^2_L(z)S_{{\rm bolo}}(1+z)^{-1},\quad E_{\rm p} = E^{{\rm obs}}_{{\rm p}}(1+z), \end{equation}
where $E^{{\rm obs}}_{{\rm p}}$ and  $S_{{\rm bolo}}$ are the GRB spectral peak energy and bolometric fluence, which are the observables.

We determine the parameters of the Amati relation from the GRB sample at
$z<1.4$, which are shown in Figure.~\ref{The apparent magnitudes reconstructed from  Gaussian processes}, by using the method of  the likelihood function
 \citep{D'Agostini2005}
\begin{eqnarray}\label{Lc}
    \mathcal{L}(\sigma,a,b,M)\propto\prod_{i=1}^{N_1} \frac{1}{\sigma^2}
    \times\exp\left[-\frac{[y_i-y(x_i,z_i; a, b, M)]^2}{2\sigma^2}\right].
\end{eqnarray}
Here $N_1=79$ or $20$  denotes  the number of low-redshift GRBs in A220 or A118 data sets,  $\sigma=\sqrt{\sigma_{\rm int}^2+\sigma_{y,i}^2+b^2\sigma_{x,i}^2}$,  $\sigma_{\rm int}$ is the intrinsic scatter of GRBs,
$\sigma_y=\frac{1}{\rm ln10}\frac{\sigma_{E_{\rm iso}}}{E_{\rm iso}},\quad \sigma_x=\frac{1}{\rm ln10}\frac{\sigma_{E_{\rm p}}}{E_{\rm p}}$, $\sigma_{E_{\rm p}}$ is the error magnitude of the spectral peak energy, and $\sigma_{E_{\rm iso}}=4\pi d^2_L\sigma_{S_{\rm bolo}}(1+z)^{-1}$ is the error magnitude of isotropic equivalent radiated energy,
where $\sigma_{S_{\rm bolo}}$ is the error magnitude of bolometric fluence.

We use the python package \texttt{emcee} \citep{ForemanMackey2013}, which is optimized on the basis of the Metropolis-Hastings algorithm, to implement Markov Chain Monte Carlo (MCMC) numerical fitting method. The absolute magnitude $M$ of SNe Ia should be fitted simultaneously  with the calibration parameters $a$ and $b$.\footnote{In the calibration procedure, it is inappropriate to directly use the distance
moduli of SNe Ia samples since the absolute magnitude $M$ is unknown.}
The number of points that have been used in each \textit{emcee} procedure is 8000.
The calibrated results are summarized in Table 1, and plotted in Figure 2.
We find that the values of absolute magnitude with the A220 GRB data set and the A118 GRB data set are almost the same ($M=-19.50^{+1.40}_{-1.40}$). The results of the intercept $a$ with the A220 GRB data set are well consistent at $1\sigma$ with the A118 GRB data set, while the difference of the slope $b$ between A220 and A118 is very significant. Moreover,  the value of the 1$\sigma$ uncertainty of the slope $b$ in A220 is smaller than that in A118, which is attributed to  the number of calibrated GRBs in A220 (79 GRBs) and this is apparently larger than the one in A118 (20 GRBs). 
Furthermore, the value of the 1$\sigma$ uncertainty of the slope $b$ in the A220 
and A118 data sets 
is smaller than  that obtained in \cite{Liu2022b}($b=1.290^{+0.126}_{-0.126}$ in A220, and $b=0.99^{+0.205}_{-0.205}$ in A118, respectively), by the linear interpolation from SNe Ia with setting $M=-19.36$ as a constant. The intrinsic scatter $\sigma_{\rm int}$ from the A118 GRB data set is smaller than the one from the A220 GRB data set. This character agrees with the results obtained in \cite{Khadka2021}, which indicate that the A118 data set is a higher-quality one compared to the A220 data set.

\setlength{\tabcolsep}{6mm}{
\begin{table*}[h]
	\centering
	\caption{Calibration Results ($a$ and $b$) of the Amati Relation at $z<1.4$ and the absolute magnitude $M$ of the Pantheon sample fitted by \textit{emcee} with A220 and A118 GRB Data Sets with  the 1$\sigma$ Uncertainty. 
}
	\label{tab:1}
	\begin{tabular}{ccccc}
		\hline\hline\noalign{\smallskip}	
		Data Sets & $a$ & $b$  &$\sigma_{\rm int}$ & $M$\\
		\noalign{\smallskip}\hline\noalign{\smallskip}
		A220 (79 GRBs) & $52.77^{+0.58}_{-0.58}$ & $1.298^{+0.090}_{-0.080}$ & $0.521^{+0.027}_{-0.034}$ & $-19.50^{+1.40}_{-1.40}$\\
		A118 (20 GRBs) & $52.93^{+0.58}_{-0.58}$ & $1.01^{+0.14}_{-0.14}$ & $0.466^{+0.046}_{-0.063}$ & $-19.50^{+1.40}_{-1.40}$\\
		\noalign{\smallskip}\hline
	\end{tabular}
\end{table*}}

\begin{figure*}[h]
\centering
\includegraphics[width=200px]{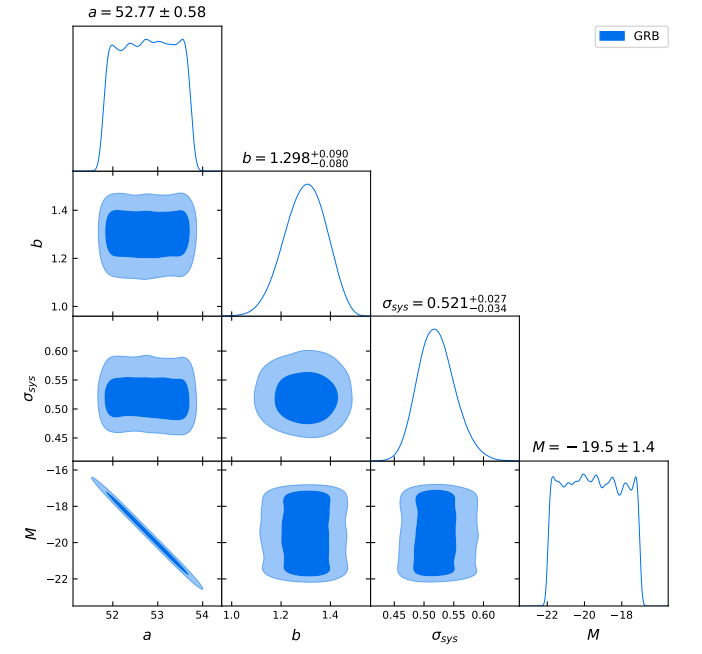}
\includegraphics[width=200px]{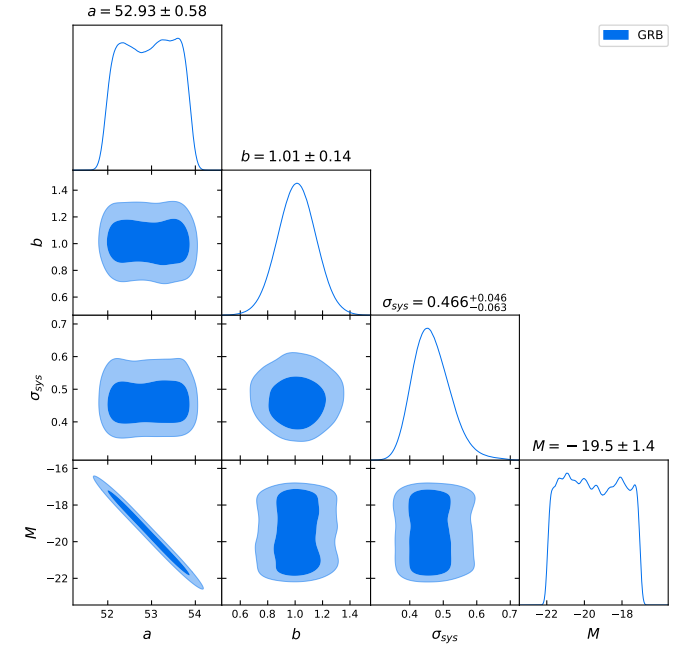}
\caption{Calibration results ($a$ and $b$) of the Amati relation at $z<1.4$ and the absolute magnitude $M$ of the Pantheon sample fitted by \textit{emcee} with GRB data A220 (left panel) and A118 (right panel), respectively.} \label{Amati relation}
\end{figure*}

\subsection{GRB Hubble diagram}
Extrapolating the results from the low-redshift GRBs to the high-redshift ones,  we are able to obtain the energy ($E_{\rm iso}$) of each burst at high redshift ($z>1.4$). Therefore, the luminosity distance ($d_{L}$) can be derived. Then, we obtain the GRB  Hubble diagram with the A219 sample\footnote{We remove one point GRB 051109A in the A220 (A102) sample \citep{Khadka2021} to obtain the A219 sample, in which 140 GRBs at $1.4<z\le8.2$ (see the Appendix for details).} and A118 sample, which are plotted in Figure 3. The derived distance moduli of 140 GRBs (A219) and 98 GRBs (A118) at $1.4<z\le8.2$ are listed in the Appendix.

The uncertainty of the GRB distance modulus with the Amati relation is
\begin{equation}\sigma^2_\mu=\bigg(\frac{5}{2}\sigma_{{\rm log}\frac{E_{\rm iso}}{\rm 1erg}}\bigg)^2+ \bigg (\frac{5}{\rm 2ln10}\frac{\sigma_{S_{\rm bolo}}}{S_{\rm bolo}} \bigg)^2 \, ,\end{equation}
where
\begin{equation}\sigma^2_{{\rm log}\frac{E_{\rm iso}}{\rm 1erg}}=\sigma^2_{\rm int}+ \bigg (\frac{b}{\rm ln10}\frac{\sigma_{E_{\rm p}}}{E_{\rm p}} \bigg )^2+\sum \bigg (\frac{\partial_{y}(x;\theta_c)}{\partial \theta_i} \bigg)^2C_{ii}\, .
\end{equation}
Here $\theta_c$=\{$\sigma_{{\rm int}}$, $a, b$\}, and $C_{ii}$ means the diagonal element of the covariance matrix of these fitting coefficients.

\begin{figure*}[h]
\centering
\includegraphics[width=200px]{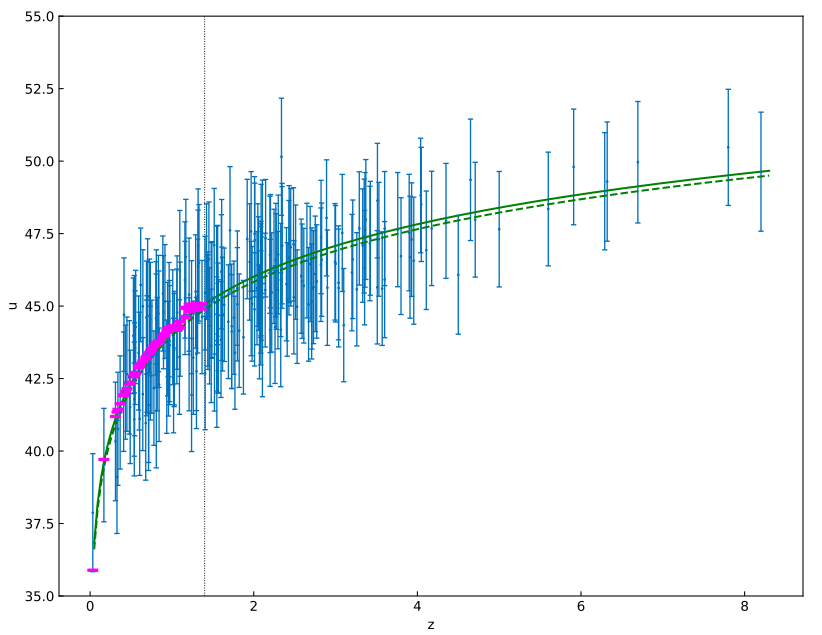}
\includegraphics[width=200px]{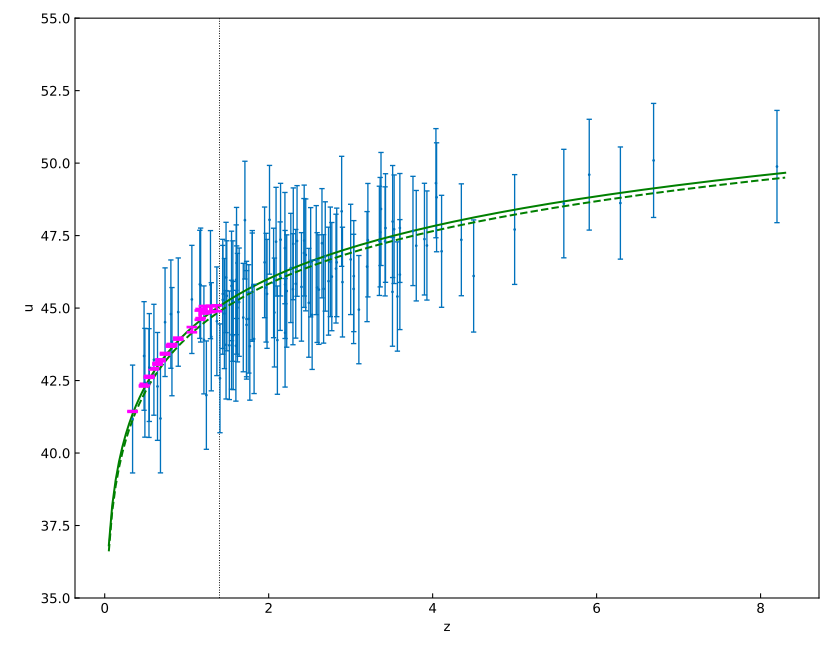}
\caption{GRB Hubble diagram with the A219 sample (left panel) and the A118 sample (right panel).  The GRBs at $z<1.4$ are obtained by Guassian process from SNe Ia data (purple dots), and the
GRBs at $z>1.4$ (blue dots) are obtained with the Amati relation calibrated with the sample at $z<1.4$. The solid green curve is the CMB standard distance modulus with $H_0=67.36\ {\rm km}\ {\rm s}^{-1}{\rm Mpc}^{-1}$, $\Omega_{\rm m}$=0.315 \citep{Plank2020}, and the green long dotted curve is the SNIa standard distance modulus with $H_0=74.3\ {\rm km}\ {\rm s}^{-1}{\rm Mpc}^{-1}$, $\Omega_{\rm m}$=0.298 \citep{Scolnic2018}. The black dashed line denotes $z=1.4$.} \label{GRB Hubble diagram}
\end{figure*}

\section{CONSTRAINTS ON COSMOLOGICAL MODELS}
When using GRB data at $z>1.4$ to constrain cosmological models, the cosmological parameters can be fitted by using the minimization $\chi^2$ method:
\begin{equation}\chi^2_{\rm GRB} = \sum^{N_2}_{i=1} \left[\frac{\mu_{\rm obs}(z_i)-\mu_{\rm th}(z_i;p,H_0)}{\sigma_{\mu_i}}\right]^2.
\end{equation}
Here $N_2=140$ or $98$  denotes  the number of high-redshift GRBs in A219 or A118 data sets,  $\mu_{{\rm th}}$ is the theoretical value of distance modulus calculated from the cosmological model,  $H_0$ is the Hubble constant,  $p$ represents the cosmological parameters, and $\mu_{{\rm obs}}$ is the observational value of distance modulus and its error $\sigma_{\mu_i}$.
Considering a flat space\footnote{The cosmological models have been usually constrained with flat spatial curvature; however, recently works constrain nonspatially flat models with GRBs and results are promising (see, e.g. \cite{Khadka2021,Cao2022a,Luongo2022}). },
for the $w$CDM model that has a constant equation of state of dark energy, the theoretical value of the luminosity distance can be obtained from
\begin{equation}d_{L;{\rm th}}=\frac{{c}(1+z)}{H_{\rm 0}}\int^z_0\frac{dz}{[\Omega_{\rm m}(1+z)^3+\Omega_{\Lambda}(1+z)^{3(1+w)}]^{\frac{1}{2}}} . \end{equation}
Here   $c$ is the speed of light, and $\Omega_{\rm m}$ and $\Omega_{\Lambda}$ are the present dimensionless density parameters of matter and dark energy, respectively, which satisfy $\Omega_{\rm m}$ + $\Omega_{\Lambda}$ = 1.
For the flat $\Lambda$CDM model, $w=-1$.

Except for the GRB data, we also include  the OHD to constrain cosmological models.  The OHD  can be obtained from the galactic age differential method \citep{Jimenez2002}, which has advantages to constrain cosmological parameters and distinguish dark energy models.  The 31 Hubble parameter measurements at $0.07<z<1.965$ \citep{Stern2010,Moresco2012,Moresco2015,Moresco2016,Zhang2014,Ratsimbazafy2017} are used in our analysis. For the OHD data set, the $\chi^2$ has the form
\begin{equation}\chi^2_{{\rm {OHD}}} =
\sum^{N_3}_{i=1} \left [\frac{H_{\rm  obs}(z_i)-H_{\rm th}(z_i;p,H_0)}{\sigma_{H_i}} \right]^2 . \end{equation}
Here $N_3=31$  denotes  the number of the Hubble parameter measurements. Thus, the total $\chi^2$ of GRB and OHD data is
\begin{equation}\chi^2_{{\rm total}} = \chi^2_{{\rm GRB}} + \chi^2_{{\rm OHD}}.
\end{equation}

We use the python package \texttt{emcee} \citep{ForemanMackey2013} to constrain cosmological models.
In each \textit{emcee} procedure we generate 8000 datasets.
The results only with 140 GRBs (A219)  and 98 GRBs (A118) are shown in Figure 4 ($\Lambda$CDM model) and Figure 5 ($w$CDM model); and the joint results from 140 GRBs (A219)  and 98 GRBs (A118) combined with 31 OHD  are shown in Figure 6 ($\Lambda$CDM model) and Figure 7 ($w$CDM model). The constraints of Figure 4-7 with the 1$\sigma$ confidence level are summarized in Table 2\footnote{GRB data
alone are unable to constrain $H_0$ because of the degeneracy between $H_0$ and the correlation intercept
parameter; therefore $H_0$ is set to be $70\ {\rm km}\ {\rm s}^{-1}{\rm Mpc}^{-1}$ for GRB-only analyses in previous works \citep{Khadka2021, Cao2022a}. In order to compare with the previous analyses, we also set $H_0$=$70\ {\rm km}\ {\rm s}^{-1}{\rm Mpc}^{-1}$ for the cases only with GRBs. For a free $H_0$ in the fitting procedure, \texttt{emcee} will provide the similar numerical outcome of the $\Omega_{\rm m}$ value,  and $H_0$ is a bound parameter if the absolute magnitude M of SNe Ia is set as a concrete value in the low-redshift calibration.}. With 98 GRBs at $1.4<z<8.2$ in the A118 sample, we obtained $\Omega_{\rm m}$ = $0.51^{+0.11}_{-0.17}$ for the flat $\Lambda$CDM model, and $\Omega_{\rm m}$ = $0.47^{+0.21}_{-0.17}$,  $w$ = $-0.98^{+0.75}_{-0.48}$ for the flat $w$CDM model.  Combining 98 GRBs in the A118 sample with 31 OHD, we obtained $\Omega_{\rm m}$ = $0.346^{+0.048}_{-0.069}$ and $h$ = $0.677^{+0.029}_{-0.029}$  for the flat $\Lambda$CDM model, and  $\Omega_{\rm m}$ = $0.314^{+0.072}_{-0.055}$, $h$ = $0.705^{+0.055}_{-0.069}$, $w$ = $-1.23^{+0.33}_{-0.64}$ for the flat $w$CDM model. Here $h\equiv \frac{H_0}{100 \mathrm{km/s/Mpc}}$. It should be noted that the $\Lambda$CDM model ($w=-1$) is consistent within 1$\sigma$ with 98 GRBs at $1.4<z<8.2$ in A118 sample and 31 OHD data sets for the flat $w$CDM model. Our results are more stringent than the previous analyses ($\Omega_{\rm m}=0.34^{+0.13}_{-0.15}$, $w=-0.86^{+0.36}_{-0.38}$  at the 2$\sigma$ confidence level), which made use of 193 GRBs combined with the SNe Ia \citep{Amati2019}. Here we must point out that only the $\Lambda$CDM and $w$CDM models are considered in our discussions since these two models have  less model parameters.  However, the dark energy models with the redshift-evolving equation of state  appear crucial because several scenarios of them  are intriguing in order to heal the $H_0$ tension\footnote{The constraint on the Hubble constant $H_0$ can be given with very high-redshift CMB data based on the $\Lambda$CDM model ($H_0=67.36\pm0.54\ {\rm km}\ {\rm s}^{-1}{\rm Mpc}^{-1}$) \citep{Plank2020}, which has a more than $~4\sigma$ deviation from the value of obtained directly from the very low-redshift SNe Ia data ($H_0=74.3\pm1.42\ {\rm km}\ {\rm s}^{-1}{\rm Mpc}^{-1}$) \citep{Riess2018}. The $H_0$ tension seems to suggest that there are potentially unknown systematic errors in observational data, or $\Lambda$CDM model used to determine the Hubble constant may be inconsistent with the present universe. Observational data at the redshift region ($2<z<1000$) are necessary to precisely identify the possible origin of $H_0$ tension.}. We will go further with the evolving dark energy models in future work and we would expect higher error bars on parameters that predict possible dark energy evolution.

 We find that the $H_0$ value 
 from GRBs at high redshift
 and OHD  ($h$ = $0.677^{+0.029}_{-0.029}$) seem to be favor to the one from the Planck cosmic microwave backgroud (CMB) observations, which is consistent with previous analyses \citep{Liu2022b}.
 We also find the $\Omega_{\rm m}$ value of our results for the flat $\Lambda$CDM model with GRBs at high redshift and OHD ($\Omega_{\rm m}$ = $0.346^{+0.048}_{-0.069}$) is consistent with the one from the Planck CMB observations ($\Omega_{\rm m}=0.3153\pm0.00073)$\citep{Plank2020} at the 1$\sigma$ confidence level.  Our result of the $H_0$ value for the flat $\Lambda$CDM model appears very well confirmed, while much less for the flat $w$CDM, which indicate that the $H_0$ tension instead still persists.

\begin{figure*}
\centering
\includegraphics[width=150px]{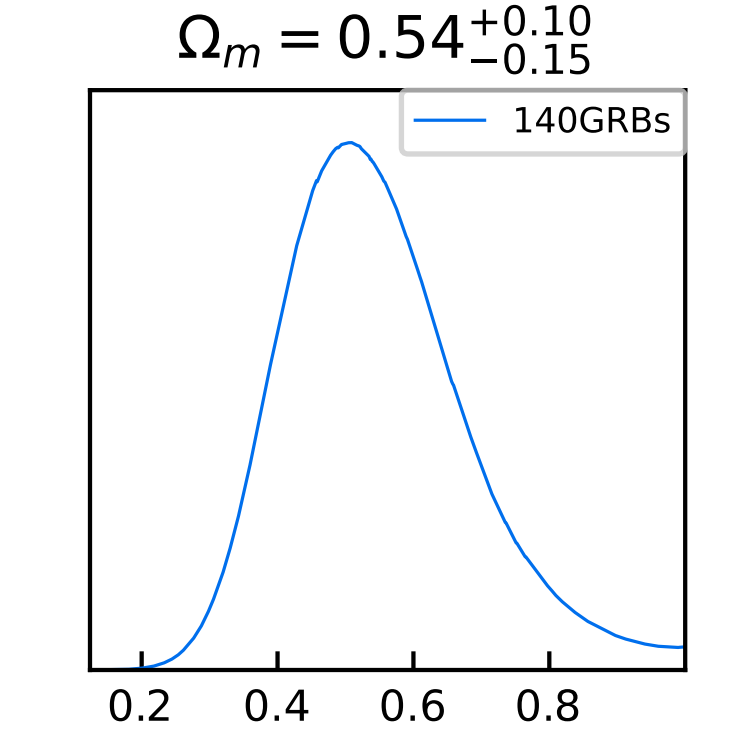}
\includegraphics[width=150px]{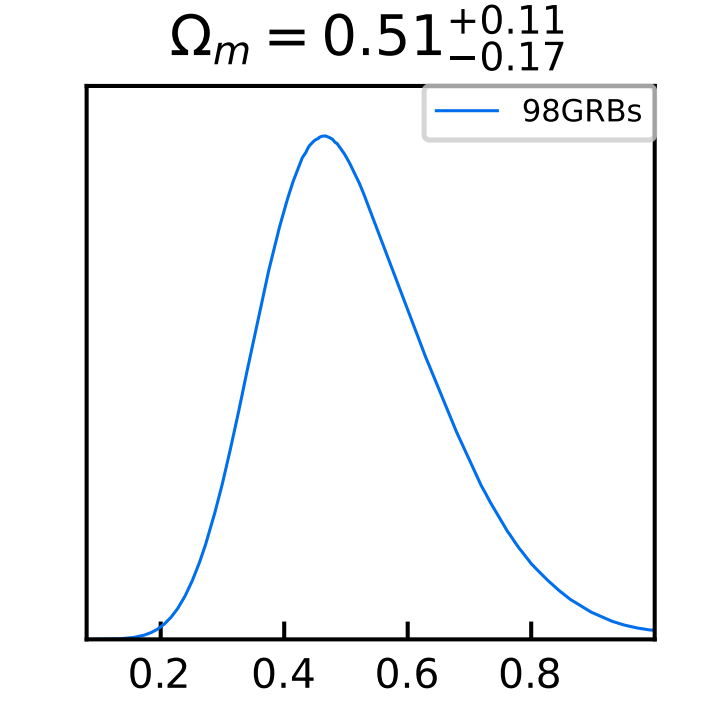}
\caption{Constraints on $\Omega_{\rm m}$ in the $\Lambda$CDM model at high redshift $z>1.4$ only with 140 GRBs (left panel), and 98 GRBs (right panel). 
$H_0$ is set to be $70\ {\rm km}\ {\rm s}^{-1}{\rm Mpc}^{-1}$.} \label{constrain}
\end{figure*}

\begin{figure*}
\centering
\includegraphics[width=160px]{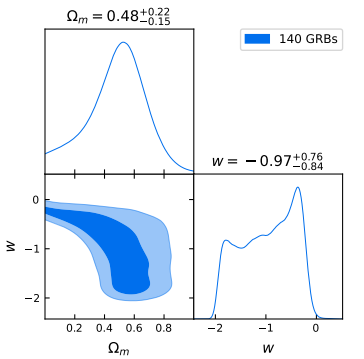}
\includegraphics[width=160px]{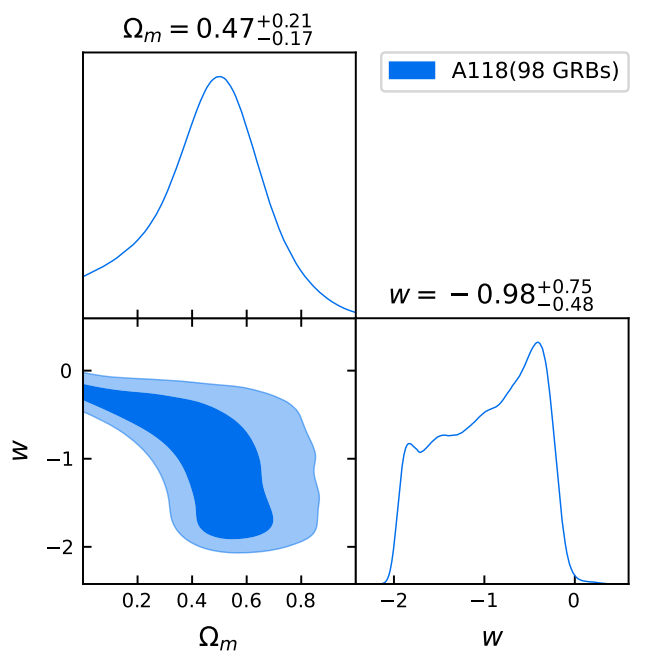}
\caption{Constraints on $\Omega_{\rm m}$ and $w$ in the $w$CDM model at high redshift $z>1.4$ only with 140 GRBs (left panel), and 98 GRBs (right panel). 
$H_0$ is set to be $70\ {\rm km}\ {\rm s}^{-1}{\rm Mpc}^{-1}$.} \label{constrain 2}
\end{figure*}

\setlength{\tabcolsep}{7mm}{
\begin{deluxetable}{clccc}
	\tablecaption{Constraints on the $\Lambda$CDM and $w$CDM Models at the 1$\sigma$ Confidence
Level from GRBs at high redshift $z>1.4$,  A219 ($z>1.4$) + 31 OHD, and A118 ($z>1.4$) + 31 OHD Data Sets. \label{tab2}}
	\tablewidth{0pt}
	\tablehead{
		Models  &$\quad$ $\quad$ Data Sets   &   $\Omega_{\rm{m}}$ &   $h$ &   $w$
	}
	\startdata
	& A219 (140 GRBs) & $0.54^{+0.10}_{-0.15}$ & - & - \\
	$\Lambda$CDM & A118 (98 GRBs) & $0.51^{+0.11}_{-0.17}$ & - & - \\
	& A219 (140 GRBs) + 31 OHD & $0.351^{+0.050}_{-0.067}$ & $0.677^{+0.029}_{-0.029}$ & - \\
	& A118 (98 GRBs) + 31 OHD & $0.346^{+0.048}_{-0.069}$ & $0.677^{+0.029}_{-0.029}$ & - \\
	\hline
	& A219 (140 GRBs) & $0.48^{+0.22}_{-0.15}$ & - & $-0.97^{+0.76}_{-0.84}$ \\
	$w$CDM & A118 (98 GRBs) & $0.47^{+0.21}_{-0.17}$ & - & $-0.98^{+0.75}_{-0.48}$ \\
	& A219 (140 GRBs) + 31 OHD & $0.336^{+0.048}_{-0.070}$ & $0.710^{+0.055}_{-0.064}$ & $-1.30^{+0.30}_{-0.59}$ \\
	& A118 (98 GRBs) + 31 OHD & $0.314^{+0.072}_{-0.055}$ & $0.705^{+0.055}_{-0.069}$ & $-1.23^{+0.33}_{-0.64}$ \\
	\enddata
	\tablecomments{For the cases only with GRBs, $h$ is set to be 0.7.}
\end{deluxetable}}

\begin{figure*}
\centering
\includegraphics[width=180px]{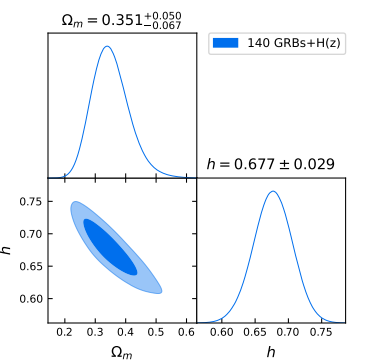}
\includegraphics[width=180px]{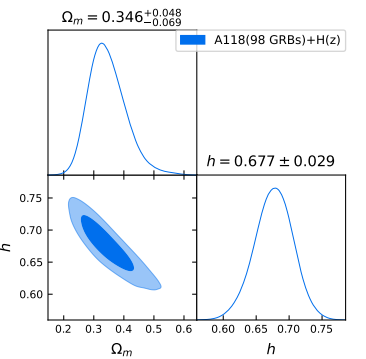}
\caption{Joint constraints on parameters of $\Omega_m$ and $h$ in the $\Lambda$CDM model at high redshift $z>1.4$ with 140 GRBs + 31 OHD (left panel), and 98 GRBs + 31 OHD (right panel).} \label{joint constrain}
\end{figure*}

\begin{figure*}
\centering
\includegraphics[width=190px]{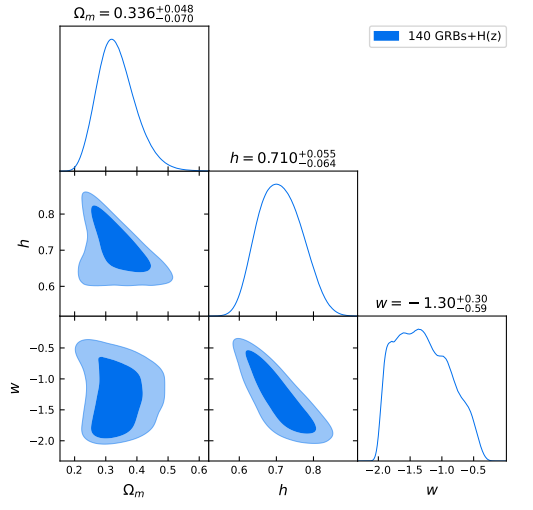}
\includegraphics[width=190px]{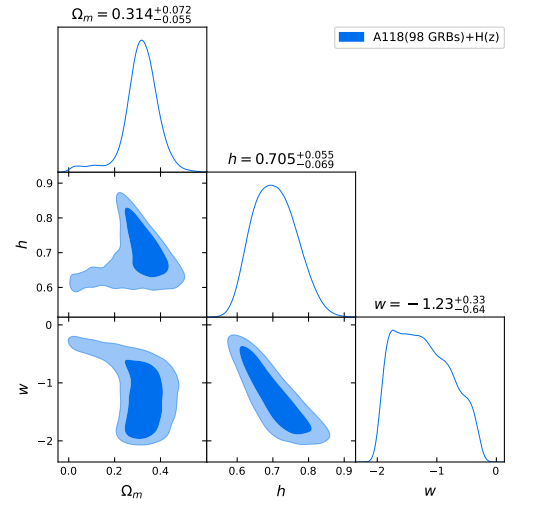}
\caption{Joint constraints on parameters of $\Omega_{\rm m}$, $h$ and $w$ in the $w$CDM model at high redshift $z>1.4$ with 140 GRBs + 31 OHD (left panel), and 98 GRBs + 31 OHD (right panel).} \label{joint constrain 2}
\end{figure*}

Finally, we also use the A219 and A118 data set to constrain the $\Lambda$CDM and $w$CDM models by using the method of simultaneous fitting. In this calculation, the parameters of cosmological models ($\Omega_{\rm m}$, $h$, and $w$) and the relation parameters ($a$ and $b$) are fitted simultaneously.
The number of points that have been used in each \textit{emcee}  procedure is 8000. The results from the A219 and A118 samples combined with the OHD data set are shown in Figure 8 ($\Lambda$CDM model) and Figure 9 ($w$CDM model), and summarized in Table 3 with the 1$\sigma$ confidence level.  With the A118 sample and 31 OHD, we obtained $\Omega_{\rm m}$ = $0.341^{+0.050}_{-0.070}$, $h$ = $0.673^{+0.029}_{-0.029}$, $a=52.984^{+0.049}_{-0.049}$, $b=1.189^{+0.084}_{-0.084}$  for the flat $\Lambda$CDM model, and  $\Omega_{\rm m}=0.325^{+0.052}_{-0.064}$, $h=0.702^{+0.048}_{-0.065}$, $w=-1.25^{+0.51}_{-0.40}$, $a=52.980^{+0.049}_{-0.049}$, $b=1.185^{+0.084}_{-0.084}$  for the flat $w$CDM model.
It is found that the values of the coefficients of the Amati relation for the flat $\Lambda$CDM model and the flat $w$CDM model are almost identical, which are consistent with the results calibrating from the low-redshift data at the 1$\sigma$ confidence level. The values of the 1$\sigma$ uncertainty of the relation parameters ($a$, $b$) and the intrinsic scatter $\sigma_{\rm int}$ in simultaneous fitting are smaller than those listed in Table 1, which is attributed to  the number of calibrated GRBs in the A220 or A118 data set, these are apparently larger than those of the 79 GRBs in the A220 sample and 20 GRBs in the A118 sample at $z<1.4$.
We also find that the simultaneous fitting results from only  GRBs are consistent with previous analyses \citep{Liu2022a}.\footnote{Our simultaneous fitting results with the A219 sample and the A118 sample are slightly different from those with the A220 sample and A118 sample obtained in \cite{Khadka2021} and \cite{Cao2022b}. It should be pointed out that there is an error in the peak energy of GRB 081121 data released in \cite{Dirirsa2019} that corresponds to the distance modulus ($47.23\pm1.08$) rather than the peak energy ($871\pm1.23$) in Table 4 of \cite{Wang2016}. Therefore, we correct this error in our work.
}
\setlength{\tabcolsep}{2mm}{
\begin{table}[!htbp]
\centering
\caption{Simultaneous Fitting Results of $\Omega_{\rm m}$, $h$, $a$, $b$ and $\sigma_{\rm int}$ in the $\Lambda$CDM and $w$CDM Models, with  A219 GRB + 31 OHD, and A118 GRB + 31 OHD Data Sets. }
\begin{tabular}{cccccccc}
\hline\hline\noalign{\smallskip}
Models& Data Sets& $\Omega_{m}$& $h$& $w$& $a$& $b$& $\sigma_{\rm int}$\\
\hline
\multirow{3}{*}{$\Lambda$CDM}&\multirow{2}{*}{A219 GRB + 31 OHD}&\multirow{2}{*}{$0.378^{+0.053}_{-0.074}$}&\multirow{2}{*}{$0.659^{+0.029}_{-0.029}$}&\multirow{2}{*}{-}&\multirow{2}{*}{$52.841^{+0.039}_{-0.039}$}&\multirow{2}{*}{$1.327^{+0.073}_{-0.073}$}&\multirow{2}{*}{$0.465^{+0.022}_{-0.025}$}\\
\multicolumn{1}{c}{\multirow{2}{*}{}}&\multirow{2}{*}{A118 GRB + 31 OHD}&\multirow{2}{*}{$0.341^{+0.050}_{-0.070}$}&\multirow{2}{*}{$0.673^{+0.029}_{-0.029}$}&\multirow{2}{*}{-}&\multirow{2}{*}{$52.984^{+0.049}_{-0.049}$}&\multirow{2}{*}{$1.189^{+0.084}_{-0.084}$}&\multirow{2}{*}{$0.392^{+0.024}_{-0.030}$}\\

\multirow{5}{*}{$w$CDM}&\multirow{4}{*}{A219 GRB + 31 OHD}&\multirow{4}{*}{$0.351^{+0.074}_{-0.067}$}&\multirow{4}{*}{$0.672^{+0.047}_{-0.064}$}&\multirow{4}{*}{$-1.14^{+0.45}_{-0.45}$}&\multirow{4}{*}{$52.842^{+0.039}_{-0.039}$}&\multirow{4}{*}{$1.330^{+0.074}_{-0.074}$}&\multirow{4}{*}{$0.463^{+0.022}_{-0.026}$}\\
\hline
\multicolumn{1}{c}{\multirow{4}{*}{}}&\multirow{4}{*}{A118 GRB + 31 OHD}&\multirow{4}{*}{$0.325^{+0.052}_{-0.064}$}&\multirow{4}{*}{$0.702^{+0.048}_{-0.065}$}&\multirow{4}{*}{$-1.25^{+0.51}_{-0.40}$}&\multirow{4}{*}{$52.980^{+0.049}_{-0.049}$}&\multirow{4}{*}{$1.185^{+0.084}_{-0.084}$}&\multirow{4}{*}{$0.392^{+0.025}_{-0.031}$}\\
\\
\\
\bottomrule
\end{tabular}
\end{table}}

\begin{figure*}
\centering
\includegraphics[width=220px]{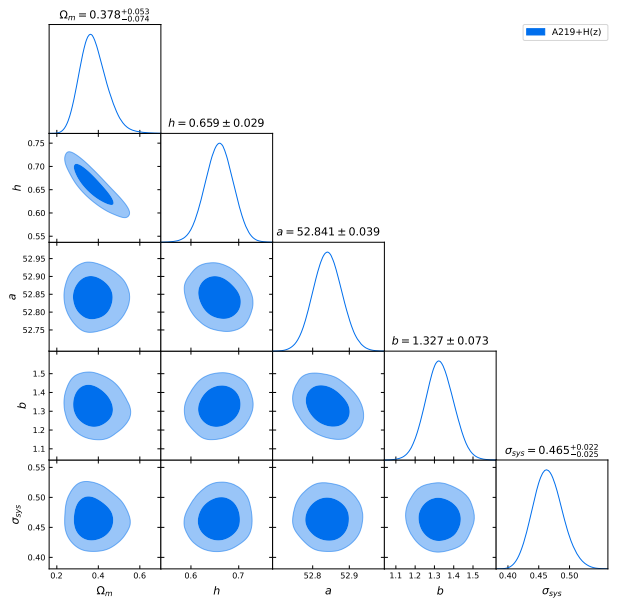}
\includegraphics[width=220px]{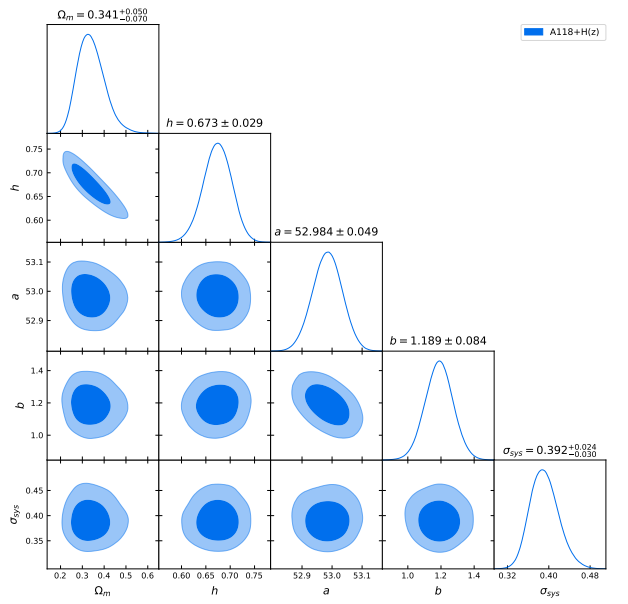}
\caption{Simultaneous fitting parameters of $\Omega_m$, $h$, $a$, $b$ and $\sigma_{int}$ in the $\Lambda$CDM model with A219 GRBs + 31 OHD (left panel), and A118 GRBs + 31 OHD (right panel). }
\label{fig/LCDM_A220_Hz.png}
\end{figure*}

\begin{figure*}
\centering
\includegraphics[width=220px]{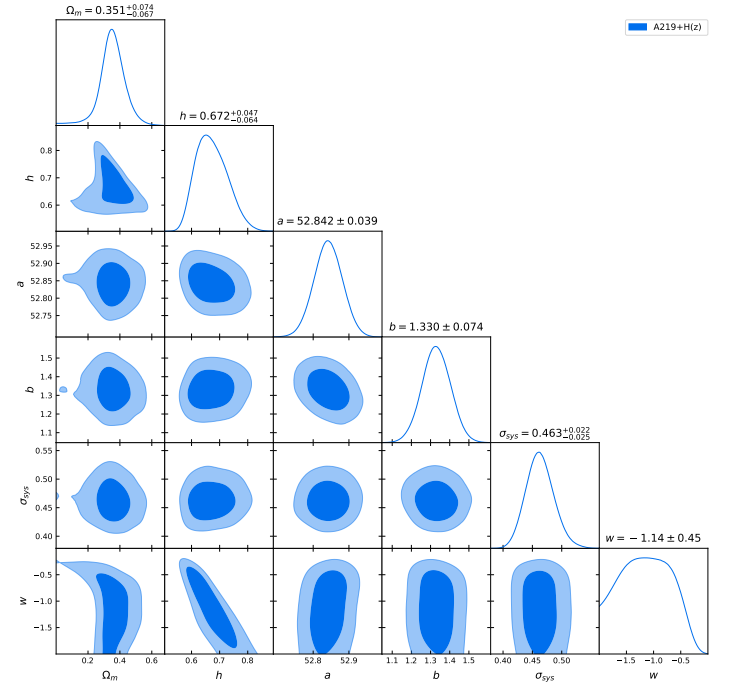}
\includegraphics[width=220px]{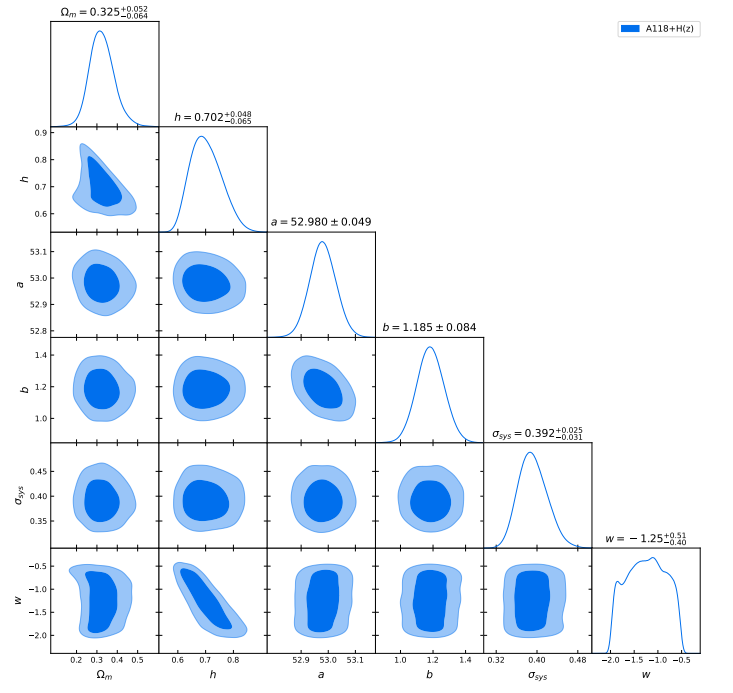}
\caption{Simultaneous fitting parameters of $\Omega_m$, $h$, $a$, $b$, $\sigma_{int}$ and $w$ in the $w$CDM model with A219 GRBs + 31 OHD (left panel), and A118 GRBs + 31 OHD (right panel).} \label{fig/LCDM_A220_Hz.png}
\end{figure*}

\section{CONCLUSIONS AND DISCUSSIONS}
In this paper, we use the Gaussian process to calibrate the Amati relation of GRB from SNe Ia data and obtain the GRB Hubble diagram with GRB data sets of the A219 and A118 samples. Then, these GRB data are used to constrain the $\Lambda$CDM and $w$CDM models.   With 98 GRBs at $1.4<z\le8.2$ in the A118 sample, we obtained $\Omega_{\rm m}$ = $0.51^{+0.11}_{-0.17}$ for the flat $\Lambda$CDM model, and $\Omega_{\rm m}$ = $0.47^{+0.21}_{-0.17}$,  $w$ = $-0.98^{+0.75}_{-0.48}$ for the flat $w$CDM model at the 1$\sigma$ confidence level.  With 98 GRBs at $1.4<z<8.2$ in the A118 sample and 31 OHD, we obtained $\Omega_{\rm m}$ = $0.346^{+0.048}_{-0.069}$ and $h$ = $0.677^{+0.029}_{-0.029}$  for the flat $\Lambda$CDM model, and  $\Omega_{\rm m}$ = $0.314^{+0.072}_{-0.055}$, $h$ = $0.705^{+0.055}_{-0.069}$, $w$ = $-1.23^{+0.33}_{-0.64}$ for the flat $w$CDM model at the 1$\sigma$ confidence level. With GRBs at high redshift and OHD date sets,  we find that the $H_0$  value seems to favor the one from the Planck CMB observations, and the $\Omega_{\rm m}$ value of our results for the flat $\Lambda$CDM model is consistent with the one from the Planck CMB observations at the 1$\sigma$ confidence level. We also use GRB data sets of A219 and A118 samples to fit $\Omega_{\rm m}$, $h$, $a$, $b$, $\sigma_{\rm int}$ and $w$ parameters simultaneously. It is found that the simultaneous fitting results are consistent with those obtained  from the low-redshift calibration method.

Furthermore, there are some discussions on possible evolutionary effects in GRB relations \citep{Li2007,BP2008,Ghirlanda2008,Tsutsui2009b,Wang2011}. Recently, \cite{Lin2016} investigated the six
relations in two redshift bins, and found moderate evidence
for the redshift evolution in four relations. \cite{Demianski2017a} found no redshift
evolution in the Amati relation with 162 GRB samples. \cite{Wang2017} found the Amati relation evolves
with redshift. More recently, \cite{Khadka2021} found that the Amati relation is independent
of redshift within the error bars with the A220 GRB data set. \cite{Dai2021} found  strong evidence that the Amati relation shows no, or marginal, evolution with redshift.
\cite{Tang2021} found that the intercept and slope of the Amati relation for the low-z subsample and high-z subsample differ at more than $2\sigma$.
As a result, whether the GRB relations are redshift dependent or not is still under debate.
Nevertheless, further examinations of possible evolutionary effects should be required for considering GRBs as standard candles for a cosmological probe.

\section*{ACKNOWLEDGMENTS}
We thank Prof. Pan Yu, Prof. Feng Jianchao, Prof. Peng Junjin, and Yang Liu, Zhang Bin, Xie Hanbei for kind help and discussions. We also thank the referee for helpful comments and constructive suggestions.
This project was supported by the Guizhou Provincail Science and Technology Foundation (QKHJC-ZK[2021] Key 020), in part by the NSFC under Grants Nos. 12275080, 12075084, 11690034, 11805063, and 12073069,   
and by the Science and Technology Innovation Plan of Hunan province under Grant No. 2017XK2019.

\appendix
\setlength{\tabcolsep}{3mm}{

\begin{deluxetable}{llc|llc|llc}\centering
	\tablecaption{List of the derived distance moduli of 140 GRBs in the A219 sample at $1.4<z\le8.2$. \label{tab4}}
	\tablewidth{0pt}
	\tablehead{
		GRB & $z$ & $\mu_{\rm{GRB}} \pm \sigma_{\mu, \rm{GRB}}$ & GRB & $z$ & $\mu_{\rm{GRB}} \pm \sigma_{\mu, \rm{GRB}}$ & GRB & $z$ & $\mu_{\rm{GRB}} \pm \sigma_{\mu, \rm{GRB}}$ 
	}
	\startdata
120711A & 1.405 & $45.44\pm1.96$ & 070521 & 2.0865 & $45.87\pm1.95$ & 050401 & 2.9 & $45.63\pm1.99$\\
160625B & 1.406 & $42.69\pm1.95$ & 150206A & 2.087 & $44.93\pm1.96$ & 141109A & 2.993 & $46.52\pm2.00$\\
151029A & 1.423 & $46.53\pm1.99$ & 061222A & 2.088 & $45.95\pm1.97$ & 090715B & 3 & $46.45\pm2.00$\\
050318 & 1.44 & $44.78\pm1.97$ & 130610 & 2.09 & $47.23\pm1.96$ & 080607 & 3.036 & $45.80\pm1.96$\\
100814 & 1.44 & $44.89\pm1.96$ & 100728B & 2.106 & $47.38\pm1.95$ & 081028 & 3.038 & $45.62\pm2.03$\\
141221A & 1.452 & $46.61\pm1.97$ & 090926A & 2.1062 & $43.82\pm1.95$ & 060607A & 3.082 & $47.52\pm2.01$\\
110213 & 1.46 & $44.31\pm2.00$ & 011211 & 2.14 & $45.55\pm1.96$ & 120922 & 3.1 & $44.34\pm1.95$\\
010222 & 1.48 & $43.62\pm1.95$ & 071020 & 2.145 & $47.34\pm2.02$ & 020124 & 3.2 & $46.15\pm2.00$\\
120724 & 1.48 & $45.19\pm1.99$ & 050922C & 2.198 & $46.76\pm2.01$ & 060526 & 3.21 & $46.61\pm2.04$\\
060418 & 1.489 & $45.22\pm1.99$ & 120624B & 2.2 & $44.18\pm1.95$ & 140423A & 3.26 & $45.57\pm1.95$\\
150301B & 1.5169 & $47.09\pm1.95$ & 121128 & 2.2 & $45.34\pm1.95$  & 140808A & 3.29 & $47.67\pm1.95$\\
030328 & 1.52 & $43.34\pm1.96$ & 080804 & 2.2045 & $46.89\pm1.95$  & 160629A & 3.332 & $47.08\pm1.95$\\
070125 & 1.547 & $44.03\pm1.96$ & 110205 & 2.22 & $45.47\pm2.05$  & 080810 & 3.35 & $47.41\pm1.96$\\
090102 & 1.547 & $45.78\pm1.97$ & 180325A & 2.248 & $46.36\pm1.97$ & 061222B & 3.355 & $46.32\pm1.97$\\
161117A & 1.549 & $42.76\pm1.95$ & 081221 & 2.26 & $44.27\pm1.94$ & 110818 & 3.36 & $47.62\pm1.97$\\
060306 & 1.559 & $45.72\pm2.05$ & 130505 & 2.27 & $46.58\pm1.95$ & 030323 & 3.37 & $47.98\pm2.06$\\
040912 & 1.563 & $44.24\pm2.24$ & 140629A & 2.275 & $46.19\pm1.99$  & 971214 & 3.42 & $47.15\pm1.97$\\
100728A & 1.567 & $43.95\pm1.95$ & 060124 & 2.296 & $45.55\pm2.02$ & 060707 & 3.425 & $47.33\pm1.96$\\
990123 & 1.6 & $43.84\pm1.99$ & 021004 & 2.3 & $46.77\pm2.05$  & 170405A & 3.51 & $45.64\pm1.95$\\
071003 & 1.604 & $46.18\pm1.96$ & 141028A & 2.33 & $45.90\pm1.95$  & 110721A & 3.512 & $48.63\pm1.97$\\
090418 & 1.608 & $46.67\pm2.00$ & 151021A & 2.33 & $44.18\pm1.96$  & 060115 & 3.53 & $47.30\pm1.96$\\
110503 & 1.61 & $45.07\pm1.95$ & 110128A & 2.339 & $50.15\pm2.01$   & 090323 & 3.57 & $45.60\pm1.95$\\
990510 & 1.619 & $44.72\pm1.96$ & 051109A & 2.346 & $47.09\pm2.02$  & 100704 & 3.6 & $47.67\pm1.96$\\
080605 & 1.6398 & $45.04\pm1.95$ & 131108A & 2.4 & $45.75\pm1.95$  & 130514 & 3.6 & $45.91\pm2.00$\\
131105A & 1.69 & $44.46\pm1.96$ & 171222A & 2.409 & $45.22\pm1.95$  & 130408 & 3.76 & $47.63\pm1.97$\\
091020 & 1.71 & $47.61\pm2.19$ & 060908 & 2.43 & $46.66\pm1.97$   & 120802 & 3.8 & $46.72\pm2.01$\\
100906 & 1.73 & $44.30\pm2.14$ & 080413 & 2.433 & $47.13\pm2.01$ & 100413 & 3.9 & $47.53\pm2.00$\\
120119 & 1.73 & $44.12\pm1.96$ & 090812 & 2.452 & $47.02\pm2.02$ & 060210 & 3.91 & $46.68\pm2.10$\\
150314A & 1.758 & $44.59\pm1.95$ & 120716A & 2.486 & $47.12\pm1.95$ & 120909 & 3.93 & $47.29\pm1.95$\\
110422 & 1.77 & $43.39\pm1.94$ & 130518A & 2.49 & $45.30\pm1.95$ & 140419A & 3.956 & $46.56\pm2.19$\\
080514B & 1.8 & $45.61\pm1.97$ & 081121 & 2.512 & $46.51\pm1.97$ & 131117A & 4.04 & $48.81\pm1.97$\\
120326 & 1.8 & $45.06\pm1.95$ & 170214A & 2.53 & $44.97\pm1.95$ & 060206 & 4.048 & $48.50\pm1.96$\\
090902B & 1.822 & $44.15\pm1.95$ & 081118 & 2.58 & $46.03\pm1.96$ & 090516 & 4.109 & $46.92\pm2.04$\\
131011A & 1.874 & $43.92\pm1.96$ & 080721 & 2.591 & $45.85\pm1.97$ &  120712A & 4.1745 & $47.68\pm1.97$\\
140623A & 1.92 & $47.31\pm2.05$ & 050820 & 2.612 & $45.70\pm1.97$ & 080916C & 4.35 & $47.93\pm1.98$\\
080319C & 1.95 & $46.52\pm2.00$ & 030429 & 2.65 & $46.57\pm1.97$ & 000131 & 4.5 & $46.07\pm2.04$\\
170113A & 1.968 & $47.57\pm2.06$ & 120811C & 2.67 & $45.06\pm1.96$ & 090205 & 4.6497 & $49.35\pm2.09$\\
081008 & 1.9685 & $45.25\pm1.97$ & 080603B & 2.69 & $46.45\pm1.98$ & 140518A & 4.707 & $47.97\pm1.97$\\
030226 & 1.98 & $45.07\pm1.97$ & 161023A & 2.708 & $45.15\pm1.98$ & 111008 & 5 & $47.65\pm1.99$\\
130612 & 2.01 & $47.49\pm1.96$ & 060714 & 2.711 & $45.60\pm2.08$ & 060927 & 5.6 & $48.34\pm1.96$\\
170705A & 2.01 & $44.88\pm1.95$& 140206A & 2.73 & $45.65\pm1.95$ & 130606 & 5.91 & $49.79\pm1.99$\\
161017A & 2.013 & $47.27\pm1.96$& 091029 & 2.752 & $46.11\pm1.99$ & 050904 & 6.29 & $48.96\pm2.02$\\
140620A & 2.04 & $45.61\pm1.95$& 081222 & 2.77 & $45.84\pm1.95$ & 140515A & 6.32 & $49.29\pm2.05$\\
081203A & 2.05 & $46.41\pm2.11$& 050603 & 2.821 & $46.42\pm1.95$ & 080913 & 6.695 & $49.96\pm2.09$\\
150403A & 2.06 & $46.12\pm1.95$& 161014A & 2.823 & $47.38\pm1.95$ & 120923A & 7.8 & $50.47\pm2.00$\\
000926 & 2.07 & $44.45\pm1.96$& 110731 & 2.83 & $46.60\pm1.95$ & 090423 & 8.2 & $49.63\pm2.05$\\
080207 & 2.0858 & $45.55\pm2.18$& 111107 & 2.89 & $48.04\pm2.00$ &  &  & \\
\enddata
\tablecomments{For the A220 sample \citep{Khadka2021}, there are two distance moduli of GRB051109A at $z=2.346$ calculated by different peak energy and bolometric fluence. In  Table 7 of \cite{Khadka2021}  for the A118 sample, $E_{\rm p}$=$539\pm200$, $S_{{\rm bolo}}$=$0.51\pm0.05$; while in Table 8 of \cite{Khadka2021} for the A102 (A220) sample, which are compiled from those listed in \cite{Demianski2017a}, $E_{\rm p}$=$538.706\pm274.372$, $S_{{\rm bolo}}$=$0.519357\pm0.269718$.
However, we find that $E_{\rm p}$=$539\pm200$, $E_{{\rm iso}}$=$6.84516\pm0.730151$ in Table 5 of \cite{Demianski2017a}, and $E_{\rm p}$=$539\pm200$, $S_{{\rm bolo}}$=$0.51\pm0.05$ in Table 1 of  \cite{Amati2008}. We remove GRB051109A in A102 (A220) sample, therefore we obtain 140 GRBs at $1.4<z\le8.2$ in the A219 sample.}
\end{deluxetable}}

\setlength{\tabcolsep}{3mm}{
\begin{deluxetable}{llc|llc|llc}\centering
	\tablecaption{List of the derived distance moduli of 98 GRBs in the A118 sample at $1.4<z\le8.2$. \label{tab5}}
	\tablewidth{0pt}
	\tablehead{
		GRB & $z$ & $\mu_{\rm{GRB}} \pm \sigma_{\mu, \rm{GRB}}$ & GRB & $z$ & $\mu_{\rm{GRB}} \pm \sigma_{\mu, \rm{GRB}}$ & GRB & $z$ & $\mu_{\rm{GRB}} \pm \sigma_{\mu, \rm{GRB}}$ 
	}
	\startdata
160625B & 1.406 & $42.57\pm1.87$& 130610 & 2.09 & $47.28\pm1.87$ & 090715B & 3 & $46.67\pm1.90$
\\
050318 & 1.44 & $45.48\pm1.88$ & 090926A & 2.1062 & $43.89\pm1.86$ & 080607 & 3.036 & $45.66\pm1.88$
\\
100814 & 1.44 & $45.28\pm1.87$ & 011211 & 2.14 & $46.10\pm1.87$ & 081028 & 3.038 & $46.10\pm1.91$
\\
110213 & 1.46 & $44.81\pm1.89$ & 071020 & 2.145 & $47.35\pm1.94$ & 120922 & 3.1 & $44.94\pm1.86$\\
010222 & 1.48 & $43.73\pm1.86$ &050922C & 2.198 & $47.06\pm1.91$ &020124 & 3.2 & $46.43\pm1.89$\\
120724 & 1.48 & $46.05\pm1.90$ &120624B & 2.2 & $44.14\pm1.87$ & 060526 & 3.21 & $47.34\pm1.95$\\	
060418 & 1.489 & $45.42\pm1.89$ &121128 & 2.2 & $45.81\pm1.86$ & 080810 & 3.35 & $47.31\pm1.88$\\	
030328 & 1.52 & $43.71\pm1.87$& 110205 & 2.22 & $45.58\pm1.93$ & 110818 & 3.36 & $47.61\pm1.88$\\
070125 & 1.547 & $44.08\pm1.87$& 130505 & 2.27 & $46.38\pm1.88$ & 030323 & 3.37 & $48.41\pm1.94$\\
090102 & 1.547 & $45.76\pm1.88$ & 060124 & 2.296 & $45.64\pm1.91$ & 971214 & 3.42 & $47.30\pm1.88$\\
040912 & 1.563 & $45.24\pm2.07$ & 021004 & 2.3 & $47.21\pm1.92$ & 060707 & 3.425 & $47.76\pm1.87$\\
100728A & 1.567 & $44.05\pm1.86$ & 141028A & 2.33 & $45.83\pm1.87$ & 170405A & 3.51 & $45.55\pm1.87$\\
990123 & 1.6 & $43.70\pm1.91$ & 051109A & 2.346 & $47.31\pm1.90$ & 110721A & 3.512 & $47.98\pm1.93$\\
071003 & 1.604 & $45.97\pm1.89$ &131108A & 2.4 & $45.73\pm1.87$ & 060115 & 3.53 & $47.71\pm1.87$\\
090418 & 1.608 & $46.55\pm1.91$ & 060908 & 2.43 & $46.90\pm1.87$ & 090323 & 3.57 & $45.39\pm1.88$
\\
110503 & 1.61 & $45.27\pm1.86$ & 080413 & 2.433 & $47.32\pm1.91$ & 100704 & 3.6 & $47.76\pm1.87$\\
990510 & 1.619 & $45.01\pm1.87$ &090812 & 2.452 & $46.83\pm1.93$ & 130514 & 3.6 & $46.15\pm1.89$\\
080605 & 1.6398 & $45.20\pm1.86$ &130518A & 2.49 & $45.18\pm1.87$ & 130408 & 3.76 & $47.65\pm1.88$\\
131105A & 1.69 & $44.67\pm1.87$ &081121 & 2.512 & $46.58\pm1.88$ & 120802 & 3.8 & $47.15\pm1.90$\\
091020 & 1.71 & $48.03\pm2.03$ & 170214A & 2.53 & $44.76\pm1.88$ & 100413 & 3.9 & $47.37\pm1.92$\\
100906 & 1.73 & $44.62\pm1.99$ & 081118 & 2.58 & $46.65\pm1.88$ & 120909 & 3.93 & $47.15\pm1.88$\\
120119 & 1.73 & $44.42\pm1.87$& 080721 & 2.591 & $45.70\pm1.89$ & 131117A & 4.04 & $49.30\pm1.88$\\
150314A & 1.758 & $44.62\pm1.86$& 050820& 2.612 & $45.63\pm1.88$ & 060206 & 4.048 & $48.82\pm1.87$\\
110422 & 1.77 & $43.68\pm1.86$ & 030429 & 2.65 & $47.23\pm1.88$ & 090516 & 4.109 & $46.95\pm1.93$\\
080514B & 1.8 & $45.78\pm1.88$& 120811C & 2.67 & $45.66\pm1.87$ & 080916C & 4.35 & $47.35\pm1.92$\\
120326 & 1.8 & $45.72\pm1.86$ & 080603B & 2.69 & $46.77\pm1.88$ & 000131 & 4.5 & $46.10\pm1.93$\\
090902B & 1.822 & $43.93\pm1.88$ & 140206A & 2.73 & $45.92\pm1.86$ & 111008 & 5 & $47.70\pm1.89$\\
080319C & 1.95 & $46.57\pm1.90$ & 091029 & 2.752 & $46.59\pm1.88$ & 060927 & 5.6 & $48.60\pm1.87$\\
081008 & 1.9685 & $45.70\pm1.87$& 081222 & 2.77 & $46.08\pm1.86$ & 130606 & 5.91 & $49.60\pm1.91$\\
030226 & 1.98 & $45.48\pm1.87$ & 050603& 2.821 & $46.36\pm1.87$ & 050904 & 6.29 & $48.62\pm1.93$\\
130612 & 2.01 & $48.04\pm1.87$& 110731 & 2.83 & $46.57\pm1.87$ & 080913 & 6.695 & $50.09\pm1.96$\\
150403A & 2.06 & $45.86\pm1.88$ & 111107& 2.89 & $48.33\pm1.89$ & 090423 & 8.2 & $49.88\pm1.93$
\\
000926 & 2.07 & $44.84\pm1.87$ & 050401 & 2.9 & $45.89\pm1.89$&&&\\
	\enddata
\end{deluxetable}}

\end{document}